\documentclass[secthm,seceqn,amsthm,ussrhead,reqno]{amsart}
\usepackage{amsmath,latexsym}
\usepackage[psamsfonts]{amssymb}
\usepackage{times}
\usepackage[mathcal]{euscript}
\numberwithin{equation}{section}

\newcommand{\bbT}{\mathbb T}

\renewcommand{\epsilon}{\varepsilon}

\newcommand{\be}{\begin{equation}}
\newcommand{\ee}{\end{equation}}
\newcommand{\no}{\nonumber}

\newcommand{\C}{\mathbb{C}}

\newcommand{\R}{\mathbb{R}}
\renewcommand{\S}{\mathbb{S}}
\newcommand{\T}{\mathbb{T}}

\newcommand{\Z}{\mathbb{Z}}

\newcommand{\cH}{{\mathcal H}}

\newcommand{\cU}{{\mathcal U}}

\renewcommand{\Im}{{\ensuremath{\mathrm{Im}}}}
\renewcommand{\Re}{{\ensuremath{\mathrm{Re}}}}

\renewcommand{\det}{\mathop{\mathrm{det}}}

{\bf}{\it}
\newtheorem{theorem}{Theorem}[section]
\newtheorem{lemma}[theorem]{Lemma}
\newtheorem{corollary}[theorem]{Corollary}
\newtheorem{assumption}[theorem]{Assumption}
\newtheorem{definition}[theorem]{Definition}
\newtheorem{proposition}[theorem]{Proposition}
\newtheorem{remark}[theorem]{Remark}

\date{\today}

\begin{document}
\title[The number of eigenvalues for an Hamiltonian in Fock space]
{The number of eigenvalues for an Hamiltonian in Fock space}

\author{Sergio  Albeverio$^{1,2,3}$, Saidakhmat  N. Lakaev$^{4,5}$,
  Tulkin H. Rasulov $^{5}$}

\address{$^1$ Institut f\"{u}r Angewandte Mathematik,
Universit\"{a}t Bonn, Wegelerstr. 6, D-53115 Bonn\ (Germany)}

\address{
$^2$ \ SFB 611, \ Bonn, \ BiBoS, Bielefeld - Bonn;}
\address{
$^3$ \ CERFIM, Locarno and Acc.ARch,USI (Switzerland) E-mail
albeverio@uni.bonn.de}

\address{
{$^4$ Samarkand State University, University Boulevard 15, 703004,
Samarkand (Uzbekistan)} \ {E-mail: slakaev@mail.ru }}

\address{
{$^5$ Samarkand State University, University Boulevard 15, 703004,
Samarkand (Uzbekistan)} \ {E-mail: rth@mail.ru }}
\begin{abstract}
A model operator $H$ corresponding to the energy operator of a
system with non-conserved number $n\leq 3$ of particles is
considered. The precise location and   structure of the essential
spectrum of $H$ is described. The existence of infinitely many
eigenvalues below the bottom of the essential spectrum of $H$ is
proved if the generalized Friedrichs model has a virtual level at
the bottom of the essential spectrum and for the number $N(z)$ of
eigenvalues  below $z<0$ an asymptotics established. The finiteness
of eigenvalues of $H$ below the bottom of the essential spectrum is
proved if the generalized Friedrichs model has a zero eigenvalue at
the bottom of its essential spectrum.
\end{abstract}
\maketitle



\section{Introduction}

One of the remarkable results in the spectral analysis for
continuous three-particle  \\Schr\"{o}dinger  operators  is the
Efimov  effect: if none of the three two-particle Schr\"{o}dinger
operators (corresponding to the two particle subsystem) has negative
eigenvalues, but at least two of them have a zero-energy resonance,
then this three-particle Schr\"{o}dinger operator has an infinite
number of discrete eigenvalues, accumulating at zero.

Since its discovery by Efimov in 1970 \cite{Efi}, many works are
devoted to this subject. See, for example,
\cite{AHW,AmNo,DFT,FaMe,OvSi,Sob,Tam91,Tam94,Yaf74}. In particular,
Yafaev gave a mathematically rigorous proof for the existence of
such a phenomenon \cite{Yaf74}.

The main result obtained by  Sobolev \cite{Sob} (see also
\cite{Tam94}) is the asymptotics of the form $\cU_0|log|\lambda||$
for the number of eigenvalues on the left of $\lambda,\lambda<0$,
where the coefficient ${\cU}_0$ does not depend on the potentials $
v_\alpha $ and is a positive function of the ratios
$m_1/m_2,m_2/m_3$ of the masses of the three-particles.

Recently  the existence of the Efimov effect for $N$-body quantum
systems with $N\geq 4$ has been proved by X.P.Wang in \cite{Wang}.

In fact in \cite{Wang} a lower bound on the number of eigenvalues of
the form $$C_0|log(E_0-\lambda)|$$ is given, when $\lambda$ tends to
$E_0$, where $C_0$ is a positive constant and $E_0$ is the bottom of
the essential spectrum.

The presence of Efimov's effect for the lattice three-particle
Schr\"odinger  operators has been proved (see, e.g.,
\cite{ALzMahp04,ALtmf03,Ltmf91,Lfa93, LAfa99,Mat,Mog91} for relevant
discussions and \cite{FIC,GrSc,KM,Mat,MS,Mog91,RSIII,Yaf00,Zh} for
the general study of the low-lying excitation spectrum for quantum
systems on lattices).

The systems considered in above mentioned works have a fixed number
of quasi-particles. In the theory of solid-state physics
\cite{Mog91} and the theory of quantum fields \cite{Frie}
interesting problems arise where the number of quasi-particles is
bounded, but not fixed. At the same time the study of systems with a
non conserved bounded number of particles is reduced to the study of
the spectral properties of self-adjoint operators acting in "the
cut" subspace ${\cH}_n$ of Fock's space, consisting of $r\le n$
particles \cite{Frie,MiSp,Mog91,MiZh}.

The perturbation problem of an operator, with point and continuous
spectrum (which acts in $\cH_2$) has played a considerable role in
the discussions about spectral problems connected with of the
quantum theory of fields (see \cite{Frie}).

The main goal of the present paper is to prove the existence of
infinitely many eigenvalues below the essential spectrum for a model
operator $H$ corresponding to the energy operators of systems of
three non conserved particles acting in the subspace ${\cH}_3.$

More precisely, under some technical smoothness assumptions  upon
the family of the generalized Friedrichs model
$h(p),\,p\in\T^3=(-\pi,\pi]^3$ (see \cite{Frie,LtsP86}) we obtain
the following results:

(i) We describe precisely  the location and structure of the
essential spectrum of $H$ by the spectrum of the generalized
Friedrichs model.

 (ii) We prove that the operator $H$
has infinitely many eigenvalues below the bottom of the essential
spectrum, if the operator $h(0)$  has a zero-energy resonance at
the bottom of its essential spectrum.

(iii) In the case (ii) we establish the following asymptotic
formula for the number $N(z)$ of eigenvalues of $H$ lying below
$z<0$
\begin{equation*}
\lim\limits_{z \to -0}\frac{N(z)}{|\log |z||}={\cU}_0 \,(0<{\cU}_0
<\infty).
\end{equation*}

(iv) We prove the finiteness of eigenvalues of $H$, if $h(0)$ has
a zero eigenvalue at the bottom of its essential spectrum.

We remark that  the presence of a zero-energy resonance for the
Schr\"{o}dinger operators is due to the two-particle interaction
operators $V$, in particular, the coupling constant (if $V$ has in
front of it a couling constant) (see,
e.g.,\cite{AGH,Ltmf92,rauch,Yaf74} ).

In our case it is remarkable that the presence of a zero-energy
resonance at the bottom of the essential spectrum of generalized
Friedrichs model (consequently the existence of the infinitely many
eigenvalues of $H$) is due to the annihilation and creation
operators acting in the symmetric Fock space.

We also notice that the assertion (iv) is also surprising and
similar assertions have not been proved for the three-particle
Schr\"odinger operators on $\R^3$ or $\Z^3$.

We remark that  the operator $H$ has been considered before, but the
existence of infinitely many eigenvalues below the bottom of the
essential spectrum of $H$ has only been announced in \cite{LRfa03}
and in the location of the essential spectrum of $H$ has been
established in \cite{LRmn03}.

The organization of present paper is as follows. Section 1 is an
introduction to the whole work. In section 2 the model operator is
described as a bounded and self-adjoint operator $H$ in the
subspace ${\cH}_3$ and the main results of the present paper are
formulated. In Section 3, we study some spectral properties of
$h(p),p\in\T^3$.
 In section 4 we describe the essential spectrum of $H.$ In Section 5,
 we reduce the eigenvalue
problem to the Birman-Schwinger principle. In section 6 we prove the
part $(i)$ of Theorem \ref{fin}. In section 7, we give an asymptotic
formula for the number of eigenvalues.
 Some technical material is collected in Appendix $A.$

Throughout the present paper we adopt the following conventions: For
each $\delta>0$ the notation $U_{\delta}(0) =\{p\in
{\bbT}^3:|p|<\delta \}$ stands for a $\delta$-neighborhood of the
origin.

Denote by  $\T^3$   the three-dimensional torus, the cube
$(-\pi,\pi]^3$ with appropriately  identified sides. Throughout the
paper the torus $\T^3$  will always be considered as an abelian
group with respect to the addition and multiplication by real
numbers regarded as operations on $\R^3$ modulo $(2\pi \Z)^3$.

\section{Three particle model operator and
statement of  results}

Let us introduce some notations used in this work. Let ${C}^1$ be
one-dimensional complex space and let $L_2({\T}^3)$ be the Hilbert
space of square-integrable functions defined on $\T^3$ and
$L_2^s(({\T}^3)^2)$ be the Hilbert space of square-integrable
symmetric functions on $({ \T}^3)^2.$

Denote by ${\cH}$ a direct sum of spaces ${\cH}_0=C^1,\,{\cH}_1=
L_2({\T}^3),\,{\cH}_2=L_2^{s}(({\T})^2),$ that is, ${\cH}= {\cH}_0
\oplus {\cH}_1 \oplus {\cH}_2.$

Let $H$ be the operator in ${\cH}$ with the entries $H_{ij}:
{\cH}_j\to {\cH}_i, i,j=0,1,2:$
$$
(H_{00}f_0)_0=u_0f_0,\quad (H_{01}f_1)_0=\int\limits_{{\T}^3}
v(q')f_1(q')dq',\quad H_{02}=0,
$$$$ H_{10}=H^*_{01},\quad
(H_{11}f_1)_1(p)=u(p)f_1(p), \quad
(H_{12}f_2)_1(p)=\int\limits_{{\T}^3} v(q')f_2(p,q')dq',
$$
$$
H_{20}=0,\quad H_{21}=H^*_{12},\quad
(H_{22}f_2)_2(p,q)=w(p,q)f_2(p,q),
$$
where $H^*_{ij}:{\cH}_i\to {\cH}_j, (j=i+1,i=0,1)$ is the adjoint
operator of $H_{ij}.$

 Here $f_i \in \cH_i,i=0,1,2,$ $ u_0$ is a real number, $v(p)$ and $u(p)$ are
 real-analytic functions on ${\T}^3$ and $w(p,q)$ is a real-analytic
symmetric function defined on $({\T}^3)^2.$

Under these assumptions the operator $H$ is bounded and
self-adjoint in ${\cH}$.

We remark that, $H_{10}$ and $H_{21}$ resp. $H_{01}$ and $H_{12}$
in the Fock space are called creation resp. annihilation
operators.

Throughout this paper we assume  the following additional
technical assumptions.

\begin{assumption}\label{hypoth1} The real-analytic function $w(p,q)$ which is
symmetric on $({\T }^3)^2,$ is even with respect to $(p,q),$ has a
unique non-degenerate zero minimum at the point $(0,0)\in
({\T}^3)^2$ and there exist positive definite matrix $ W$ and real
numbers $l_1, l_2 (l_1>0,l_2\not=0)$ such that
$$
\left( \frac{\partial^2 w(0,0)}{\partial p_i \partial p_j}
\right)_{i,j=1}^3= l_1 W,\,\, \left( \frac{\partial^2
w(0,0)}{\partial p_i \partial q_j} \right)_{i,j=1}^3= l_2 W.
$$
\end{assumption}

\begin{assumption}\label{hypoth2} The real-analytic functions $u(p)$
and $v(p)$ on ${\T}^3$ are even and the function $u(p)$ has a unique
non-degenerate minimum at $0\in {\T}^3$.
\end{assumption}

By Assumptions \ref{hypoth1} and \ref{hypoth2} for any $p\in T^3$
the integral
\begin{equation*}
\int\limits_{{\T}^3} \frac{v^2(t)dt}{w(p,t)}
\end{equation*}
is finite and hence we can define continuous function on $\T^3,$
which will be denotes  $\Lambda(p).$

Since the function $w(p,q)$ has  a unique non degenerate minimum at
the point $(0,0)\in ({\T}^3)^2$ and $w(0,0)=0$ the function
$\Lambda(p)$ is positive. In particular, if $v(0)=0$ then
$\Lambda(p)$ is a twice continuously differentiable function at the
point $p=0$ (see proof of Lemma \ref{examp})
\begin{assumption}\label{hypoth3}
(i) For any nonzero $p\in \T^3$ the inequality
$\Lambda(p)<\Lambda(0)$ holds.\\
(ii) If $v(0)=0,$ then $\Lambda(p)$ has a non-degenerate maximum
at $p=0.$
\end{assumption}
\begin{remark}
Let
$$ u(p)= \varepsilon (p)+c,\, v(p)=\varepsilon(p),\,w(p,q)=\varepsilon (p)+
\varepsilon (p+q)+ \varepsilon (q),
$$
where $c>0-$ is a real number and
\begin{align}\label{eps}
\varepsilon (p)=3-cos p_1-cos p_2 -cos p_3,\,p=(p_1,p_2,p_3) \in
{\T}^3.
\end{align}
 Then Assumptions \ref{hypoth1}, \ref{hypoth2} and \ref{hypoth3} are
fulfilled (see Lemma \ref{examp} below).
\end{remark}

To formulate the main results we introduce a family of the
generalized Friedrichs model $h(p),p\in \T^3,$ which acts in $C^1
\oplus L_2(\T^3)$ with the entries
\begin{align}\label{h}
(h_{00}(p)f_0)_0=u(p)f_0,\,\,h_{01}=\frac{1}{\sqrt{2}}H_{01},\\
h_{10}=h^*_{01}, \,\, (h_{11}(p)f_1)_1(q)=w_p(q)f_1(q),\no
\end{align}
 where $w_p(q)=w(p,q).$

Let ${\C}$ be the field of complex numbers.

The proof of the following variant of the Birman-Schwinger
principle for the family $h(p),p \in \T^3$ is customary.
\begin{proposition}\label{b-sh}
For any $p \in \T^3$ the number $z\in \C\setminus [m_w(p),M_w(p)]$
is an eigenvalue of $h(p),p \in \T^3$ if and only if the number
$1$ is an eigenvalue of the integral operator given by
$$
(\mathrm{G}(p,z)\psi)(q)=\frac{v(q)}{2(u(p)-z)}
\int\limits_{{\T}^3} \frac{v(t)\psi(t)dt}{w_p(t)-z},\,\,\psi \in
{L_2(\T^3)},
$$
where the numbers $m_w(p)$ and $M_w(p)$ are defined by
\begin{equation*}
m_w(p)=\min_{q\in {\T}^3}w(p,q) \,\, \mbox{and}\,\,
M_w(p)=\max_{q\in {\T}^3} w(p,q).
\end{equation*}
\end{proposition}

Let $C(\T^3)$ be the Banach space of continuous functions on
$\T^3.$

\begin{assumption}\label{hypoth4} For any
$p\in \T^3$ and $\psi\in C(\T^3)$ the integral
$$\int\limits_{{\T}^3} \frac{v(t)\psi(t)dt}{w_p(t)-m_w(p)}$$
is finite.
 \end{assumption}

\begin{remark}
Let $v$ be an arbitrary analytic function on $\T^3$ and
$$  w_p(q)=l_1\varepsilon (p)+
l_2\varepsilon (p+q)+ l_1\varepsilon (q),
$$ where the function $\varepsilon(p)$ is defined by
\eqref{eps} and $l_1,l_2>0,\,l_1\neq l_2$.

Then Assumption \ref{hypoth4} is fulfilled (see Lemma \ref{int
finite} below).
\end{remark}

It should be noted that under Assumption \ref{hypoth4} for any
$p\in\T^3$ the operator $G(p,z)$ can be defined as a bounded
operator on $C(\T^3)$ even for $z=m_w(p).$

\begin{definition}\label{resonance0}
Let Assumption \ref{hypoth4} be fulfilled. The operator $h(p),p \in
\T^3,$ is said to have a resonance at the point $z=m_w(p)$ if the
number  $1$ is an eigenvalue of the integral operator given by
$$
(\mathrm{G}(p,m_w(p))\psi)(q)=\frac{v(q)}{2(u(p)-m_w(p))}
\int\limits_{{\T}^3} \frac{v(t)\psi(t)dt}{w_p(t)-m_w(p)},\,\,\psi
\in {C(\T^3)},
$$ and  $v(p)\neq 0$ (provided $u(p)-m_w(p)\neq 0).$
\end{definition}

Let us denote by $\tau_{ess}(H)$ the bottom of the essential
spectrum of the operator $H$ and by $N(z)$ the number of
eigenvalues of $H$ lying below $z \leq \tau_{ess}(H).$

The main result of this work is the following.

\begin{theorem}\label{fin} (i) Assume Assumptions \ref{hypoth1},\ref{hypoth2}
and \ref{hypoth3} are fulfilled and let the operator $h(0)$ have a
zero eigenvalue. Then the operator $H$ has a finite
number of negative eigenvalues.\\
 (ii) Assume Assumptions \ref{hypoth1}, \ref{hypoth2} and part (i) of Assumption
 \ref{hypoth3}
 are fulfilled and the operator $h(0)$ has a
zero-energy resonance.
  Then the operator $H$ has
infinitely many negative eigenvalues accumulating at
$\tau_{ess}(H)=0$ and the function $N(z)$
 obeys the relation
\begin{equation} \label{asym.K}
\lim\limits_{z \to -0}\frac{N(z)}{|\log |z||}={\cU}_0 \,(0<{\cU}_0
<\infty).
\end{equation}
\end{theorem}
\begin{remark} The constant ${\cU}_0$ does not depend on the
function $v$  and is given as a positive function depending only on
the ratio $\frac {l_1}{l_2}$ (with $l_1,\,l_2$ as in Assumption
\ref{hypoth1}).
\end{remark}
\begin{remark}
A zero-energy resonance (resp. a zero eigenvalue) of $h(0),$ if it
does exist, is simple (see Lemma \ref{simple resonans} below).
\end{remark}
\begin{remark}
We remark that if the Assumptions \ref{hypoth1}, \ref{hypoth2} and
part (i) of Assumption \ref{hypoth3} are fulfilled then
$\tau_{ess}(H)=0.$
\end{remark}

\begin{remark}\label{techn}
We note that the assumptions for the functions $u,\,v$ and $w$ are
far from the precise, but we will not develop this point here.
\end{remark}

\section{Spectral properties of the operators $h(p),p\in\T^3$}

In this section we study  some spectral properties of  the family of
generalized Friedrichs model $h(p),\,\,p \in \T^3$ given by
\eqref{h}, which plays a crucial role in the  study of the spectral
properties of $H$. We notice that the spectrum and resonances of the
generalized Friedrichs model have been studied in detail in
\cite{ALaams97,LtsP86}.

Let the operator $h_0(p)$ act in $C^1 \oplus L_2(\T^3)$ as
$$
h_0(p) \left( \begin{array}{ll}
f_0\\
f_1(q)
\end{array} \right)=
\left( \begin{array}{ll}
u(p)f_0\\
w_p(q)f_1(q)
\end{array} \right).
$$
The perturbation $h(p)-h_0(p)$ of the operator $h_0(p)$ is a
two-dimensional bounded self-adjoint operator. Therefore in
accordance with the invariance of the absolutely continuous
spectrum under the trace class perturbations the absolutely
continuous spectrum of $h(p)$ fills the following interval on the
real axis:
$$
\sigma_{ac}(h(p))=[m_w(p),M_w(p)].
$$

For any $p \in \T^3$ and $z{\in } {\C} { \setminus }
\sigma_{ac}(h(p))$ we define the function
\begin{equation*}\label{det}
\Delta(p,z)=u(p)-z-\frac{1}{2} \int\limits_{{\T}^3}
\frac{v^2(t)dt}{w_p(t)-z}.
\end{equation*}

Note that  ${\Delta}( p, z)$ is real-analytic in $\T^3\times ({\C}
{ \setminus } \sigma_{ac}(h(p))).$

\begin{lemma}\label{delta=0}
For all $p\in \T^3$ the operator $h(p)$ has an eigenvalue $z \in
{\C} \setminus \sigma_{ac}(h(p))$ outside of the absolutely
continuous spectrum if and only if $\Delta(p,z)=0.$
\end{lemma}
\begin{proof}
The number $z \in {\C} \setminus \sigma_{ac}(h(p))$ is an
eigenvalue of $h(p)$ if and only if
 (by the Proposition \ref{b-sh})
$\lambda=1$ is an eigenvalue of the operator $G(p,z).$ According
to Fredholm's theorem  the number $\lambda=1$ is an eigenvalue for
the operator $G(p,z)$ if and only if
$$
u(p)-z-\frac{1}{2}\int\limits_{{\T}^3} \frac{v^2(t)dt}{w_p(t)-z}
=0, \quad \mbox{that is,}\quad \Delta(p,z)=0.
$$
\end{proof}

Since the function $\Delta(0,\cdot)$ is decreasing on $(-\infty,0)$
and the function $w_0(q)$ has a unique non-degenerate minimum at
$q=0$ (see Lemma \ref{minimum}) by dominated convergence the finite
limit
$$
\Delta(0,0)=\lim_{z\to -0} \Delta(0,z)
$$
exists.
\begin{lemma}\label{h0 resonans}
Let Assumption \ref{hypoth1}  be fulfilled. The operator $h(0)$
has a zero-energy resonance if and only if $\Delta(0,0)=0$ and
$v(0)\not=0.$
\end{lemma}
\begin{proof}
"Only If Part".  Let the operator $h(0)$ have a zero energy
resonance. Then by
 Definition \ref{resonance0} the equation
\begin{equation*}
\psi(q)= \frac{v(q)}{2u(0)} \int\limits_{{\T}^3}
\frac{v(t)\psi(t)dt}{w_0(t)},\,\,\psi\in {C(\T^3)},
 \end{equation*}
 has a simple solution $\psi \in C({\T^{3}})$.

One can  check that this solution is equal to the function $v(q)$
(up to a constant factor).
 Therefore we see that
 $$
v(q)=\frac{v(q)}{2u(0)} \int\limits_{{\T}^3}
\frac{v^2(t)dt}{w_0(t)}
 $$
and hence
$$ \Delta(0,0)= u(0)-\frac{1}{2}
\int\limits_{{\T}^3}\frac{v^2(t)dt}{w_0(t)} =0.
$$
"If Part". Let  the equality $ \Delta(0,0)= 0$  hold and $v(0)\neq
0$. Then  only the function $v(q) \in  C({\T}^{3})$ is a solution of
the equation
 $$
\psi(q)= \frac{v(q)}{2u(0)} \int\limits_{{\T}^3}
\frac{v(t)\psi(t)dt}{w_0(t)},$$
  that is, the
operator $h(0)$ has a zero energy resonance.
\end{proof}

\begin{lemma}\label{zeroeigen} Let Assumption \ref{hypoth1}
 be fulfilled.
  The  operator $h(0)$ has a zero eigenvalue if and only if
$\Delta(0,0)=0$ and  $v(0)=0.$
\end{lemma}
\begin{proof}
"Only If Part". Suppose $f=(f_0,f_1),$ is an eigenvector of the
operator $h(0)$ associated with the zero eigenvalue. Then
$f_0,f_1(q)$ satisfy the system of equations
\begin{equation}\label{sistema}
\left \lbrace
\begin{array}{llll}
u(0)f_0+ \frac{1}{\sqrt{2}}\int\limits_{{\T}^3} v(q')f_1(q')dq'=0\\
\frac{1}{\sqrt{2}}v(q)f_0+w_0(q)f_1(q)=0.
\end{array} \right.
\end{equation}
From \eqref{sistema} we find that $f_0$ and $f_1(q),$ except for an
arbitrary factor, are given by
\begin{equation}\label{eigenunction}
f_0=1,\,f_1(q)=-\frac{v(q)}{\sqrt{2}w_0(q)},
\end{equation}
and from the first equation of (\ref{sistema}) we derive the
equality
$$
\Delta(0,0)=0.
$$
Since the functions $w_0(q)$ and $v(q)$ are analytic on $\T^3$ and
the function $w_0(q)$ has a unique non-degenerate minimum at the
origin we can conclude  that $f_1\in L_2(\T^3)$ if and only if
$v(0)=0.$

"If Part". Let $v(0)=0$ and  $\Delta(0,0)=0$ then the vector
$f=(f_0,f_1),$ where
$$
f_0=const\neq 0,\,f_1(q)=-\frac{v(q)}{2u(0)w_0(q)}f_0,
$$
obeys the equation
$$ h(0)f=0$$
and
$$
f_1\in L_2(\T^3).
$$
\end{proof}

\begin{lemma}\label{simple resonans}
Let Assumption \ref{hypoth1} be fulfilled and the operator $h(0)$
have a zero-energy resonance (resp. zero eigenvalue). Then the
vector $f=(f_0,f_1),$ where $ f_0$ and $f_1$ are given by
\eqref{eigenunction}, is the unique solution (up to a constant
factor) of the equation $h(0)f=0$ and $f_1\in L_1(\T^3)\setminus
L_2(\T^3)$ (resp. $\,f_1\in L_2(\T^3)).$
\end{lemma}

\begin{proof}
Let the operator $h(0)$ have a zero-energy resonance (resp. a zero
eigenvalue). Then by Lemma \ref{h0 resonans} (resp. Lemma
\ref{zeroeigen}) we have $ v(0)\neq 0$ (resp. $v(0)=0)$ and
$\Delta(0,0)=0.$

One can check that  $f=(f_0,f_1)$ obeys the equation $h(0)f=0$ or
the system of equations (\ref{sistema}).

Since the functions $w_0(q)$ and $v(q)$ are analytic on $\T^3$ and
the function $w_0(q)$ has a unique non-degenerate minimum at the
origin we can conclude that $f_1\in L_1(\T^3)\setminus L_2(\T^3)$
(resp. $f_1\in L_2(\T^3))$ if and only if $v(0)\neq 0$ (resp.
$v(0)=0).$

From the representation \eqref{eigenunction}  of $f_0$ and $f_1$
 it follows that the subspace generated by the vector $f=(f_0,f_1)$ is
one dimensional.
\end{proof}

\begin{lemma}\label{neg eigen} Let Assumption \ref{hypoth1} be fulfilled.

(i) Let $\max\limits_{p\in\T^3}\Delta(p,0)<0.$ Then for any $p \in
\T^3$ the operator $h(p)$ has a unique negative eigenvalue.\\
(ii) Let $\min\limits_{p\in\T^3}\Delta(p,0)<0$ and
$\max\limits_{p\in\T^3}\Delta(p,0)\ge 0.$ Then there exists a non
void open set $D\subset \T^3$ such that for any $p \in D$ the
operator $h(p)$ has a unique negative eigenvalue and for $p \in
\T^3\setminus D$  the operator $h(p)$ has no negative
eigenvalues.\\
(iii) Let $\min\limits_{p\in\T^3}\Delta(p,0)\ge 0.$ Then for any
$p \in \T^3$ the operator $h(p)$ has no negative eigenvalues.
\end{lemma}

\begin{proof}$(i)$ Let $\max\limits_{p\in\T^3}\Delta(p,0)<0.$ Since $\T^3$ is a compact
set and the function $\Delta(p,0)$ is continuous on $\T^3$ for all
$p \in \T^3$ we have the inequality
\begin{equation*}
\Delta(p,0)<0.
\end{equation*}
For any $p\in\T^3$ the function $\Delta(p,\cdot)$ is continuous
and decreasing on $(-\infty,0]$ and
$$
\lim_{z\to -\infty} \Delta(p,z)=+\infty.
$$
Then there exist a unique point $z(p)\in (-\infty,0)$ such that
$\Delta(p,z(p))=0.$ Hence by Lemma \ref{delta=0} for any $p\in\T^3$
the point $z(p)$ is the unique negative eigenvalue of $h(p),\,p \in
\T^3.$

 $(ii)$ Let $\min\limits_{p\in\T^3}\Delta(p,0)<0$ and
$\max\limits_{p\in\T^3}\Delta(p,0)\ge 0.$

Let us introduce the notation
\begin{equation*}
D\equiv \{p\in \T^3: \Delta(p,0)<0\}.
\end{equation*}

Since the function $\Delta(p,0)$ is continuous on $\T^3$ and $\T^3$
is compact there exist points $p_0,p_1\in\T^3$  such that
$$
\min\limits_{p\in\T^3}\Delta(p,0)=\Delta(p_0,0)<0,\,
\max\limits_{p\in\T^3}\Delta(p,0)=\Delta(p_1,0)\ge 0
$$
and we have that $D\neq \T^3$ is a non void open set.

If $p\in D,$ then $\Delta(p,0)<0$ and it is proved as above that for
any $p \in D,$ the operator $h(p)$ has a unique negative eigenvalue.

Since the function $\Delta(p,\cdot)$ is decreasing on $(-\infty,0]$
for all $ p \in \T^3\setminus D$ and $z<0,$ we have
\begin{equation*}
\Delta(p,z)>\Delta(p,0)\ge 0.
\end{equation*}

 Then by Lemma \ref{delta=0}  for all $p \in \T^3\setminus D$  the operator
$h(p)$ has no negative eigenvalues.

$(iii)$ Let $\min\limits_{p\in\T^3}\Delta(p,0)\ge 0.$ Since $\T^3$
is a compact set and the function $\Delta(p,0)$ is continuous on
$\T^3$ we have
\begin{equation*}
\Delta(p,0)\ge 0\quad \mbox{for all}\quad p\in \T^3
\end{equation*}
and it is proved as above that for all $p\in \T^3,$ the operator
$h(p)$ has no  negative eigenvalues.
\end{proof}

The following decomposition plays a crucial role in the proof of
the asymptotics  \eqref{asym.K}.
 \begin{lemma}\label{razlojeniya} Assume Assumptions \ref{hypoth1}
 and \ref{hypoth2} are fulfilled.
 Then for any $p\in U_{\delta}(0),\delta>0$ sufficiently small, and
 $z\leq 0$  the following decomposition holds:
\begin{align}\label{raz}
\Delta(p,z)=
 \Delta(0,0)+
2 \pi^2 v^2(0) \frac{\sqrt{l_1^2-l_2^2}}{l_1^2} (det W)^{-
\frac{1}{2}}\sqrt{m_w(p)-z}+
\\+\Delta^{(02)}(m_w(p)-z)+
\Delta^{(20)}(p,z),\nonumber
\end{align}
where $\Delta^{(02)}(m_w(p)-z)$ (resp. $\Delta^{(20)}(p,z)$) is a
function behaving like $O({m_w(p)-z})$ (resp. $O(|p|^2)$) as
 $|{m_w(p)-z}| \to 0$ (resp. $p\to 0$).
\end{lemma}

\begin{proof}
 Let $W(p,q)$ the function defined on
$U_{\delta}(0)\times \T^3$  as
 \begin{equation}\label{w}
 W(p,q)=w_p(q+q_0(p))-m_w(p),
\end{equation}
where $q_0(p)\in\T^3$ is an analytic function in $p\in
U_{\delta}(0)$ (see Lemma \ref{minimum}) and   is the
non-degenerate minimum point of the function $w_p(q)$ for any
$p\in U_{\delta}(0).$

 We define the function
$\tilde\Delta(p,\zeta)$ on $ U_\delta(0)\times {\C}_{+}$  by
 $$
\tilde\Delta(p,\zeta)=\Delta(p,m_w(p)-\zeta^2),$$ where
${\C}_{+}=\{z\in\C: \Re z>0\}$. Using \eqref{w} the function
$\tilde\Delta(p,\zeta)$ is represented as
\begin{equation*}
\tilde {\Delta}( p, \zeta) = u(p)-m_w(p)+\zeta^2-\frac{1}{2}
\int\limits_{{\T}^3} \frac{v^2(q+q_0(p))}{W(p,q)+\zeta^2 }dq.
\end{equation*}

Let $V_{\delta}(0)$ be a complex $\delta$-neighborhood of the
point $\zeta=0 \in \C$. Denote by $\Delta^{*}(p,\zeta)$ the
analytic continuation of the function $\tilde\Delta(p,\zeta)$ to
the region $U_\delta(0)\times ({\C}_{+} \cup V_{\delta}(0))$.
Since the functions $v(q),u(q),m_w(q)$ and $W(p,q)$
 are even we have that $\Delta^{*}(p,\zeta)$   is even in $p\in U_\delta(0).$
Then by the asymptotics $u(p)=u(0)+O(|p|^2)$ as $p\to 0$ we have
\begin{equation}
\label{raz1}
 \Delta^{*}(p,\zeta)=\Delta^{*}(0,\zeta)+
 \tilde\Delta^{(20)}(p,\zeta),
\end{equation}
 where $\tilde\Delta^{(20)}(p,\zeta)=O(|p|^2)$
 uniformly in  $\zeta \in {\C_{+}}$ as $p\to 0$ (see also \cite{Ltmf92}).
A Taylor series expansion gives
 \begin{equation}
\label{raz2} \Delta^{*}(0,\zeta)=\Delta^{*}(0,0)+
\tilde\triangle^{(01)}(0,0)\zeta+
\tilde\triangle^{(02)}(0,\zeta)\zeta^2,
\end{equation} where
$\tilde\triangle^{(02)}(0,\zeta)=O(1)\quad\text{as}\quad \zeta\to
0.$

 A simple computation shows that
\begin{equation}\label{partial}
\frac{\partial \Delta^{*}(0,0)}{\partial
w}=\tilde\triangle^{(01)}(0,0)= 2 \pi^2 v^2(0)
\frac{\sqrt{l_1^2-l_2^2}}{l_1^2} (det W)^{- \frac{1}{2}} .
\end{equation}

The representations \eqref{raz1} , \eqref{raz2} and the equality
\eqref{partial} give \eqref{raz}.

\end{proof}

\begin{corollary}\label{razl.lemma.natijasi.} Assume Assumptions \ref{hypoth1}
 and \ref{hypoth2} are fulfilled and let the operator $h(0)$ have a zero energy
 resonance.  Then for any $p\in U_{\delta}(0),\delta>0$ sufficiently small, and
 $z\leq 0$  the following decomposition holds:
\begin{align*}
\Delta(p,z)= 2 \pi^2 v^2(0) \frac{\sqrt{l_1^2-l_2^2}}{l_1^2} (det
W)^{- \frac{1}{2}}\sqrt{m_w(p)-z}+
\\+\Delta^{(02)}(m_w(p)-z)+
\Delta^{(20)}(p,z),\nonumber
\end{align*}
where the functions $\Delta^{(02)}(m_w(p)-z)$ and
$\Delta^{(20)}(p,z)$ are the same as in Lemma \ref{raz}.
\end{corollary}

\begin{proof}
The proof of Corollary \ref{razl.lemma.natijasi.} immediately
follows from decomposition \eqref{raz} and Lemma \ref{h0
resonans}.
\end{proof}
\begin{corollary}\label{razl.lemma.natijasi.0}
 Assume Assumptions
\ref{hypoth1} and \ref{hypoth2} are fulfilled and let the operator
$h(0)$ have a zero energy resonance.
 Then there exists $\delta>0$  such that for any $p\in U_{\delta}(0),\,p\neq 0 $
\begin{equation}\label{Lamb.ineq}
\Delta(p,0)>0,\quad \mbox{that\,\, is,}\quad \Lambda(p)<\Lambda(0).
\end{equation}
\end{corollary}
\begin{proof}
By  Corollary \ref{razl.lemma.natijasi.} and the asymptotics (see
part (ii) of Lemma \ref{minimum})

\begin{equation}\label{malfa}
m_w(p)= \frac{l^2_1-l^2_2}{2l_1}(Wp,p)+O(p^4)\quad \mbox{as} \quad
p\to 0
\end{equation}
 we get
$$
2 \pi^2 v^2(0) \frac{\sqrt{l_1^2-l_2^2}}{l_1^2} (det W)^{-
\frac{1}{2}}\sqrt{m_w(p)}> |\Delta^{(20)}(p,0)|
$$
for  $p\in U_\delta(0),p\neq 0,\delta>0-$ sufficiently small. This
inequality gives \eqref{Lamb.ineq}.
\end{proof}

\begin{lemma}\label{D.ineq}Assume Assumptions \ref{hypoth1},\ref{hypoth2}
 and \ref{hypoth3} are fulfilled and let the operator $h(0)$ have a zero-energy
 resonance. Then there exist positive numbers $c,C$ and $\delta$ such that
\begin{equation}\label{c<(.,.)<c}
c |p| \leq \Delta(p,0 ) \leq C |p|\quad\mbox{for any}\quad p\in
U_\delta(0)
 \end{equation}
and
\begin{equation}\label{(.,.)>c}
\Delta(p,0 ) \geq c  \quad\mbox{for any}\quad\mbox p\in
\T^3\setminus U_\delta(0).
 \end{equation}
\end{lemma}
\begin{proof}
From \eqref{raz} and \eqref{malfa} we get \eqref{c<(.,.)<c} for some
positive numbers $c,C$.

By Assumptions \ref{hypoth2} and \ref{hypoth3}  we get
$\Delta(p,0)>0,\, p\neq 0$. Since $\Delta(p,0 )$ is continuous on
$\T^3$ and $\Delta(0,0 )=0$ we have \eqref{(.,.)>c} for some
$c>0.$
\end{proof}
 \begin{lemma}\label{main.ineq} Assume Assumptions \ref{hypoth1},\ref{hypoth2}
 and \ref{hypoth3} are fulfilled and let the operator $h(0)$ have a zero eigenvalue,
 then there exist  numbers $\delta>0$ and $c>0$ so that
\begin{align*}
|\Delta(p,0)|\geq c p^2 \quad \mbox{for all}\quad p\in
U_\delta(0),\\ |\Delta(p,0)|\geq c
 \quad \mbox{for all}\quad
p\in \T^3\setminus U_\delta(0).
\end{align*}
\end{lemma}
\begin{proof} Let the operator $h(0)$ have a zero
eigenvalue. By Lemma \ref{zeroeigen} we have $\Delta(0,0)=0$ and
$v(0)=0.$ By Assumptions  \ref{hypoth2} and \ref{hypoth3} the
function $\Delta(p,0)=u(p)-\frac{1}{2}\Lambda(p)$ has a unique
non-degenerate minimum at $p=0$. Then there exist positive numbers
$\delta$ and $c$ such that the statement of the lemma is
fulfilled.
\end{proof}

\section{The essential spectrum of the operator $H$}

We consider the operator $\hat{H}$ acting in
$\hat{\cH}=L_2({\T}^3)\oplus L_2(({\T}^3)^2)$ as
$$
\hat{H} \left( \begin{array}{ll}
f_1(p)\\
f_2(p,q)
\end{array} \right)=
\left( \begin{array}{ll}
u(p)f_1(p)+ \frac{1}{\sqrt{2}} \int\limits_{{\T}^3} v(q')f_2(p,q')dq'\\
\frac{1}{\sqrt{2}}v(q)f_1(p)+w_p(q)f_2(p,q)
\end{array} \right).
$$
The operator $\hat{H}$ commutes with any multiplication operator
$U_\gamma$ by the function $\gamma (p)$ acting in  $\hat{\cH}$ as
$$
U_\gamma \left(
\begin{array}{ll}
f_1(p)\\
f_2(p,q)
\end{array} \right)=
\left( \begin{array}{ll}
\gamma(p)f_1(p)\\
\gamma(p)f_2(p,q)
\end{array} \right), \gamma \in L_2({\T}^3).
$$
Therefore the decomposition of the space $\hat{\cH}$ into the
direct integral
$$
\hat{\cH}= \int\limits_{{\T}^3} \oplus \hat{\cH}(p)dp,
$$
where $\hat{\cH}(p)= C^1 \oplus L_2({\T}^3),$ yields for the
operator $\hat{H}$ the  decomposition into the direct integral
\begin{equation}\label{decompose}
 \hat{H}= \int\limits_{{\T}^3} \oplus h(p)dp,
 \end{equation}
where we recall that the fiber operators $h(p),p\in\T^3,$ are
defined by (\ref{h}).

\subsection{The spectrum of the operator $\hat H$  }

\begin{lemma}\label{spec}
For the spectrum  $\sigma(\hat H)$  of  $\hat H$ the equality
$$
\sigma(\hat H)\equiv\cup_{p\in {\T}^3} \sigma_d(h(p))\cup [0, M]
$$
holds, where $\sigma_d(h(p))$ is the discrete spectrum of
$h(p),p\in\T^3$.
\end{lemma}
\begin{proof} The assertion of the lemma follows from the representation \eqref{decompose}
of the operator $\hat H$ and the theorem on decomposable operators
(see \cite{RSIV}).
\end{proof}

Set
\begin{align}\label{two.branch}
&\sigma_{two}(\hat H)=\cup_{p\in {\T}^3} \sigma_d(h(p)),\\
&a\equiv\inf \sigma_{two}(\hat H),\,\,b\equiv\sup\sigma_{two}(\hat
H).\nonumber
\end{align}

So by Lemma \ref{delta=0} the operator $h(p),p\in \T^3,$ has in the
interval $(M,+\infty)$ either one or zero number of eigenvalues.
Hence the location and structure of the essential spectrum of $\hat
H$ can be precisely described as well as in the following

\begin{lemma}\label{two.bran.lem} Assume Assumption \ref{hypoth1} is
fulfilled and let $\Delta(p,M)\le 0$ for any $p\in \T^3$.

 (i) Let $\max\limits_{p\in\T^3}\Delta(p,0)<0,$ then
\begin{equation*} \sigma(\hat H)= [a,b]\cup [0,
M]\quad \mbox{and}\quad b<0.
\end{equation*}
(ii) Let $\min\limits_{p\in\T^3}\Delta(p,0)<0$ and
$\max\limits_{p\in\T^3}\Delta(p,0)\ge 0,$ then
\begin{equation*}
\sigma(\hat H)= [a, M]\quad \mbox{and}\quad a<0.
\end{equation*}
(iii) Let $\min\limits_{p\in\T^3}\Delta(p,0)\ge 0,$ then
$$
\sigma(\hat H)= [0,M]. $$
\end{lemma}

\begin{proof}
$(i)$. Let $\max\limits_{p\in\T^3}\Delta(p,0)<0$. Then by Lemma
\ref{neg eigen} for all $p\in \T^3$ the operator $h(p)$ has a unique
negative  eigenvalue $z(p)<m_w(p)$.

By Assumptions \ref{hypoth1} and \ref{hypoth2} and Lemma
\ref{delta=0} $z: p\in  \T^3 \to  z(p)$ is a real analytic
function on $\T^3.$

Therefore $\Im z$ is a connected  closed subset of $(-\infty,0)$,
that is, $\Im z=[a,b]$ and $b<0$ and hence $\sigma_{two}(\hat
H)=[a,b].$

$(ii)$. Let $\min\limits_{p\in\T^3}\Delta(p,0)<0$ and
$\max\limits_{p\in\T^3}\Delta(p,0)\ge 0$. Then by assertion $(ii)$
of Lemma \ref{neg eigen} there exists a non void  open set $D$ such
that for any $p\in D $ the operator $h(p)$ has a unique negative
eigenvalue $z(p)$.

Since for any $p\in \T^3$ the operator $h(p)$ is bounded and $\T^3$
is compact set, there exist a positive number $C$ such that
$\sup\limits_{p\in \T^3}||h(p)||\le C$ and for any $p\in \T^3$ we
have
\begin{equation}\label{spectrsubset}
\sigma(h(p))\subset [-C,C].
\end{equation}

For any $q\in \partial D=\{p\in \T^3: \Delta(p,0)=0\}$ there exist
$\{p_n\}\subset D$ such that $p_n\to q$ as $n\to \infty.$ Set
$z_n=z(p_n).$ Then by Lemma \ref{neg eigen} for any $p_n\in D$ the
number $z_n$ is negative and from \eqref{spectrsubset} we get
$\{z_n\}\subset [-C,0].$ Without loss of a generality we assume that
$\{z_{n}\}\to z_0$ as $n\to \infty.$

The function $\Delta(p,z)$ is continuous in $\T^3\times
(-\infty,0]$ and hence
$$
0=\lim\limits_{n\to \infty}\Delta(p_{n},z_{n}) =\Delta(q,z_0).
$$

Since for any $p\in \T^3$ the function $\Delta(p,\cdot)$ is
decreasing  in $(-\infty,0]$ and $p\in
\partial D$ we can see that $\Delta(p,z_0)=0$ if and only if
$z_0=0.$

For any $q\in \partial D$ we define
$$
z(q)=\lim\limits_{p\to q, p\in D}z(p) =0.
$$

Since the function $z(p)$ is continuous on the compact set
$D\cup\partial D$ and $z(p)=0,\,\,p\in
\partial D$ and we conclude that
$$
\Im z=[a,0],\quad a<0.
$$

Hence the set
$$
\{z \in \sigma_{two}(\hat H): z\le0\}= \cup_{p\in \T^3}
\sigma_d(h(p))\cap (-\infty,0]
$$
coincides with the set $\Im z=[a,0].$ Then Lemma \ref{spec} and
\eqref{two.branch} complete the proof of $(ii).$

$(iii)$. Let $\min\limits_{p\in\T^3}\Delta(p,0)\ge 0$. Then by Lemma
\ref{neg eigen} for all $p\in \T^3$ the operator $h(p)$ has no
negative eigenvalues.

Hence we have
\begin{equation*}
 \sigma(\hat H)=[0,M].
\end{equation*}
\end{proof}

\begin{lemma}\label{ess.spec.H} The essential spectrum
$\sigma_{ess}(H)$ of the operator $H$ coincides with the spectrum
of $\hat H,$ that is,
\begin{equation}\label{essH}
 \sigma_{ess}(H)=\sigma(\hat H).
\end{equation}
\end{lemma}
\begin{proof}
In \cite{LRmn03} it has been proved that the essential spectrum
$\sigma_{ess}(H)$ of the operator $H$ coincides with
$\sigma_{two}(\hat H)\cup [0,M].$ By Lemma \ref{spec} we have
\eqref{essH}.
\end{proof}

From Lemmas \ref{two.bran.lem} and \ref{ess.spec.H} we have the
following theorem.

\begin{theorem}\label{ess.loc} Assume Assumption \ref{hypoth1} is
fulfilled and let $\Delta(p,M)\le 0$ for any $p\in \T^3$.

(i) Let $\max\limits_{p\in\T^3}\Delta(p,0)<0,$ then
\begin{equation*} \sigma_{ess}(H)= [a,b]\cup [0,
M]\quad \mbox{and}\quad b<0.
\end{equation*}
(ii) Let $\min\limits_{p\in\T^3}\Delta(p,0)<0$ and
$\max\limits_{p\in\T^3}\Delta(p,0)\ge 0,$ then
\begin{equation*}
\sigma_{ess}(H)= [a, M]\quad \mbox{and}\quad a<0.
\end{equation*}
(iii) Let $\min\limits_{p\in\T^3}\Delta(p,0)\ge 0,$ then
$$
\sigma_{ess}(H)= [0,M]. $$
\end{theorem}

\section{The Birman-Schwinger principle}
In this section we prove an analogue of the Birman-Schwinger
principle.

Let $M(z),z<\tau_{ess}(H),\,z\neq u_0$ the operator in $\cH$ with
entries
$$
M_{00}(z)=M_{11}(z)=M_{22}(z)=M_{02}(z)=M_{20}(z)=0,
$$
otherwise
$$
M_{\alpha\beta}(z)=-R_\alpha^{\frac{1}{2}}(z) H_{\alpha\beta}
R_\beta^{\frac{1}{2}}(z),
$$
where $R_{\alpha}(z)=(H_{\alpha\alpha}-z)^{-1},\,\alpha=0,1,2.$
\begin{proposition}\label{MzVz}
The number $\lambda>1$ is an eigenvalue of the operator
$M(z),z<\tau_{ess}(H),\\z\neq u_0$ if and only if the number
$\lambda^2$ is an eigenvalue of the operator in $L_2(\T^3)$ given by
$$
V(z)=R_1^{\frac{1}{2}}(z)H_{10}R_0(z)H_{10}R_1^{\frac{1}{2}}(z)
+R_1^{\frac{1}{2}}(z)H_{12}R_2(z)H_{21}R_1^{\frac{1}{2}}(z).
$$
Moreover the eigenvalues $\lambda$ and $\lambda^2$ have the same
multiplicities.
\end{proposition}
\begin{proof}
Let $\lambda>1$ be an eigenvalue of  $M(z),$ that is,  the
equation $M(z)f=\lambda f$ or a system of equations
\begin{equation}\label{MZL}
\left \lbrace
\begin{array}{llll}
-M_{01}(z)f_1=\lambda f_0\\
-M_{10}(z)f_0-M_{12}(z)f_2=\lambda f_1\\
-M_{21}(z)f_1=\lambda f_2
\end{array} \right.
 \end{equation}
has a nontrivial solutions.

From the first and third equations of the system (\ref{MZL}) for
$f_\alpha,\alpha=0,2$ we get
$$
f_\alpha=-\frac{1}{2}M_{\alpha 1}(z)f_1,\,\alpha=0,2.
$$
Substituting the latter expression for $f_\alpha,\alpha=0,2$ into
the second equation of the system (\ref{MZL}) we have the
following
\begin{align}\label{VZL}
V(z)f_1=\lambda^2 f_1,\, f_1\in L_2(\T^3)
\end{align}
and  this equation has a nontrivial solution if and only if the
system of equations \eqref{VZL} has a nontrivial solution and
 the linear subspaces of solutions of
\eqref{MZL} and \eqref{VZL} have the same dimensions.
\end{proof}

In our analysis of the spectrum of $H$ the crucial is role played
by the compact integral operator  $ T(z), z<\tau_{ess}(H),\,z\neq
u_0$ in the space $L_2({\T}^3) $ with the kernel
$$
\frac{v(p)v(q)}{\sqrt{\Delta(p,z)}\sqrt{\Delta(q,z)}} \left
\lbrack \frac{1}{2(w(p,q)-z)}+ \frac{1}{u_0-z} \right \rbrack.
$$

For a bounded self-adjoint operator $A,$ we define $n(\lambda,A)$
as
$$
n(\lambda,A)=sup\{ dim F: (Au,u) > \lambda,\, u\in F,\,||u||=1\}.
$$
$n(\lambda,A)$ is equal the infinity if $\lambda$ is in essential
spectrum of $A$ and if $n(\lambda,A)$ is finite, it is equal to the
number of the eigenvalues of $A$ bigger than $\lambda$. By the
definition of $N(z)$ we have
$$ N(z)=n(-z,-H),\,-z
> -\tau_{ess}(H).
$$

The following lemma is a realization of well known
Birman-Schwinger principle for the operator $H$ (see
\cite{ALzMahp04,Sob,Tam94}).
\begin{lemma}\label{b-s}
The operator $T(z),z\neq u_0$ is compact and continuous in
$z<\tau_{ess}(H)$ and
\begin{equation}\label{N=n}
N(z)=n(1,T(z)).
\end{equation}
\end{lemma}

\begin{proof}
Since
$$
H=\left(
\begin{array}{ccc}
H_{00} & 0 & 0\\
0 & H_{11} & 0\\
0 & 0 & H_{22}\\
\end{array}
\right)+  \left(
\begin{array}{ccc}
0 & H_{01} & 0\\
H_{10} & 0 & H_{12}\\
0 & H_{21} & 0\\
\end{array}
\right),
$$
and $H_{ii}-z,i=0,1,2$ is positive and invertible for
$z<\tau_{ess}(H),z\neq u_0$ (for simplicity we assume $z>u_0,$ if
then $z<u_0$ the operator $-(H_{00}-z)$ is positive) one has $f \in
\cH$ and $((H-z {\bf I})f,f)<0$ if and only if $((M(z)-{\bf
I})v,v)>0$ and $v_i=(H_{ii}-z)^{\frac{1}{2}}f_i,i=0,1,2,$ where
${\bf I}$ is the identity operator on $\cH.$

It follows that
\begin{equation}\label{N(z)=}
N(z)=n(1,M(z)).
\end{equation}
Using Proposition \ref{MzVz} we get
\begin{equation}\label{n(1,M(z))=}
n(1,M(z))=n(1,V(z)).
\end{equation}

Now we represent the operator $H_{21}$ as a sum of two operators
$H^{(1)}_{21}$ and $H^{(2)}_{21}$ acting from $L_2(\T^3)$ to
$L_2((\T^3)^2)$ as
$$
(H_{21}^{(1)}f_1)(p,q)=\frac{1}{2} v(p)
f_1(q),\,(H_{21}^{(2)}f_1)(p,q)= \frac{1}{2} v(q)f_1(p).
$$

Then $\varphi\in L_2(\T^3)$ and $((V(z)-I)\varphi,\varphi)>0$ iff
$\psi=R_1^{\frac{1}{2}}(z)\varphi$ and
\begin{equation*}
((H_{11}-z-H_{12}R_2(z)H^{(2)}_{21})\psi,\psi)<
(H_{10}R_0(z)H_{10}\psi,\psi)+(H_{12}R_2(z)H^{(1)}_{21}\psi,\psi),
\end{equation*}
where ${I}$ is the identity operator on $L_2(\T^3).$ This fact means
that
\begin{equation}\label{n(1,V(z))=}
n(1,V(z))=n(-z, H_{11}-H_{12}R_2(z)H^{(2)}_{21}-H_{10}R_0(z)H_{01}
-H_{12}R_2(z)H^{(1)}_{21}).
\end{equation}

Since $z<\tau_{ess}(H),$   for any $p\in \T^3$ the function
${\Delta}( p, z)$ is positive and hence the operator
$H_{11}-z-H_{12}R_2(z)H^{(2)}_{21}$ is positive and invertible and
$$
(H_{11}-z-H_{12}R_2(z)H^{(2)}_{21})^{-\frac{1}{2}}=R_{11}^{\frac{1}{2}}(z)>0.
$$

A direct calculation shows that
\begin{equation}\label{n(1,T(z))=}
n(-z, H_{11}-H_{12}R_2(z)H^{(2)}_{21}-H_{10}R_0(z)H_{01}
-H_{12}R_2(z)H^{(1)}_{21}=n(1,T(z)).
\end{equation}
The equalities \eqref{N(z)=},\eqref{n(1,M(z))=},
\eqref{n(1,V(z))=} and \eqref{n(1,T(z))=} gives \eqref{N=n}.

Finally it we note that the operator $T(z),z\neq u_0$ is compact and
continuous in $z<\tau_{ess}(H).$
\end{proof}

\section{ The finiteness of the  number of eigenvalues of the  operator $H$.}

\begin {lemma}\label{G-S} Assume Assumptions \ref{hypoth1},\ref{hypoth2}
 and \ref{hypoth3} are fulfilled and let the operator $h(0)$ have a zero eigenvalue.
 Then  the operator $T(z),\,z\neq u_0,\,(u_0\neq 0)$ belongs to the Hilbert-Schmidt
 class and
is continuous from the left up to $z=0$.
\end{lemma}
\begin{proof}
Since the function $v(p)$ is analytic, even and $v(0)=0$ we have
$|v(p)|\leq C|p|^2$ for some $C>0$.
 By virtue of Lemmas \ref{D.ineq}, \ref{main.ineq}  and
\ref{U.ineq} the kernel of the operator $T(z),\,z\leq 0,\,z\neq
u_0,\,(u_0\neq 0)$ is estimated by
$$ C\Big (
 \frac{\chi_{\delta}(p)}{|p|}+1)(
\frac{|p|^2|q|^2 \chi_{\delta}(p)\chi_{\delta}(q)}{p^2+q^2}+1
 )(\frac{\chi_{\delta}(q)}{|q|}+1)
 \Big),
$$ where  $ \chi_{\delta}(p)$ is the characteristic function of
$U_\delta(0).$

Since the latter function is square integrable on $(\T^3)^2$ we have
that operator $T(z),\,z\neq u_0$ is a Hilbert-Schmidt operator.

The kernel function of $T(z),\,z\neq u_0$ is continuous in $p,q \in
\T^3,\,z<0$ and square integrable
 on $(\T^3)^2$ as $z\leq 0$.
Then by the dominated convergence theorem the operator $T(z)$ is
continuous from the left up to $z=0.$
\end{proof}

We are now ready for the

 {\bf Proof of $(i)$ of Theorem \ref{fin}.} Let the conditions
 part $(i)$ of Theorem \ref{fin} be fulfilled.

{\bf Case $u_0\neq 0.$} By Lemma \ref{b-s}  we have
$$ N(z)=n(1,T(z)),\,z\neq u_0\,\mbox{as}\,\,z<0 $$ and by Lemma \ref{G-S}
for any $\gamma\in [0,1)$ the number $n(1-\gamma,T(0)) $ is finite.
Then we have
$$
n(1,T(z))\leq n(1-\gamma,T(0))+n(\gamma,T(z)-T(0))
$$
for all $z<0,\,z\neq u_0$ and $\gamma \in (0,1).$ This relation can
be easily obtained by use of the Weyl inequality
$$
n(\lambda_1+\lambda_2,A_1+A_2)\leq n(\lambda_1,A_1)+n(\lambda_2,A_2)
$$
for sum of compact operators $A_1$ and $A_2$ and for any positive
numbers $\lambda_1$ and $\lambda_2.$

Since $T(z),\,z\neq u_0$ is continuous from the left up to $z=0$, we
obtain $$ \lim_{z\to 0} N(z)= N(0)\leq n(1-\gamma,T(0))\,\,
\mbox{for all}\,\, \gamma \in (0,1). $$

Thus
 $$N(0)\leq n(1-\gamma,T(0))<\infty.$$ The latter
inequality  proves the assertion $(i)$ of Theorem \ref{fin} in case
$u_0\neq 0$.

{\bf Case $u_0=0.$} First we represent the operator $T(z),z<0$ as
a sum of two bounded operators $T^{(1)}(z),z<0$ and
$T^{(2)}(z),z<0$ acting on $L_2(\T^3)$ as
$$
(T^{(1)}(z)f)(p)= \frac{v(p)}{2
\sqrt{\Delta(p,z)}}\int\limits_{\T^3}
\frac{v(q)f(q)dq}{\sqrt{\Delta(q,z)}(w(p,q)-z)},
$$
$$
(T^{(2)}(z)f)(p)=-\frac{v(p)}{z \sqrt{\Delta(p,z)}}
\int\limits_{\T^3}\frac{v(q)f(q)dq}{\sqrt{\Delta(q,z)}}.
$$

We remark that the range of the operator $T^{(2)}(z)$ is one and
hence $n(1,T^{(2)}(z))\leq 1$ for any $z<0.$
 Then according to Lemma \ref{b-s} and the
 Weyl inequality, for all
$z<0$ and $\gamma \in (0,1)$we obtain
 $$
N(z)=n(1,T(z))\leq n(1-\gamma,T^{(1)}(z))+1.
 $$
Since $T^{(1)}(z)$ is continuous from the left up to $z=0$, we
obtain $ N(0)\leq n(1-\gamma,T(0))$ for all $\gamma \in (0,1).$ The
latter inequality  proves the assertion $(i)$ of Theorem \ref{fin}
in case $u_0=0$.\qed

\section{Asymptotics for the number of
eigenvalues of the  operator $H$.}

In this section we shall closely follow  A. Sobolev's method
\cite{Sob} to derive the asymptotics for the number of eigenvalues
of $H$.

We shall first establish the asymptotics of $n(1,T(z)),\,z\neq u_0$ as $z\to
-0.$ Then Theorem \ref{fin} will be deduced by a perturbation
argument based on the following lemma.

 \begin{lemma}\label{comp.pert}
 Let $A (z)=A_0 (z)+A_1 (z),$ where $A_0(z)$ $(A_1(z))$ is
compact and continuous in $z<0$ $(z\leq 0).$  Assume that for some
function $f(\cdot),\,\, f(z)\to 0,\,\, z\to -0$ the limit $$
\lim_{z\rightarrow -0}f(z)n(\lambda,A_0 (z))=l(\lambda), $$ exists
and is continuous in $\lambda>0.$ Then the same limit exists for
$A(z)$ and $$ \lim_{z\rightarrow -0}f(z)n(\lambda,A
(z))=l(\lambda). $$
\end{lemma}
For the proof of Lemma \ref{comp.pert}, see Lemma 4.9 of \cite{Sob}.

By Assumption \ref{hypoth1} we get
\begin{equation}\label{asymp1}
w(p,q)=\frac{1}{2}\big(
l_1(Wp,p)+2l_2(Wp,q)+l_1(Wq,q))+O(|p|^4+|q|^4)\,\, \text{as}\,\,
p,q\rightarrow 0.
\end{equation}
By the \eqref{malfa} and Corollary \ref{razl.lemma.natijasi.} for
any sufficiently small negative $z$ we get
\begin{equation}\label{asymp2}
\Delta(p,z) = \frac{4\pi^2 b^2(0)}
 {l_1^{{3}/{2}} \det(W)^{\frac{1}{2}}}
 \left [ n (Wp,p) -2z \right ]^{\frac{1}{2}}+
 O(|p|^2+|z|)\,\, \text{as}\,\, p,z\rightarrow 0, \end{equation}
where
\begin{equation*}
n={(l^2_1-l^2_2)}/{l_1}.
\end{equation*}

 Let $T(\delta;|z|),\,z\neq u_0\neq 0$ be an
operator on $L_2({\T}^3)$ defined by
\begin{align*}
(T(\delta;|z|)f)(p)= \mathrm{d}_0 \int\limits_{\T^3} \frac{
\hat \chi_\delta (p) \hat \chi_\delta (q) (n(Wp,p)+ 2|z|)^{-1/4}
(n(Wq,q)+ 2|z|)^ {-1/4} } {l_1(Wp,p)+ 2l_2(Wp,q)+l_1(Wq,q)
 +2|z|} f(q)d q,
\end{align*}
where $\hat \chi_\delta(\cdot)$ is the  characteristic function of
$\hat U_\delta(0)=\{ p\in \T^3:\,\, |W^{\frac{1}{2}}p|<\delta \}$
and
$$
\mathrm{d}_0= \frac{{\det W}^{\frac{1}{2}}} {2
\pi^2}l_1^{\frac{3}{2}}.
$$

The main technical point to apply Lemma \ref{comp.pert} is the
following
\begin{lemma}\label{H-Sh} Let Assumptions \ref{hypoth1}, \ref{hypoth2} and part (i) of
Assumption \ref{hypoth3} be fulfilled and $u_0\neq 0$. Then the
operator $ T(z)-T(\delta; |z|),\,z\neq u_0$ belongs to the
Hilbert-Schmidt class and is continuous in
 $z\leq 0.$
\end{lemma}
\begin{proof}
Applying the asymptotics \eqref{asymp1},\eqref{asymp2} and Lemmas
\ref{D.ineq} and \ref{U.ineq} one can estimate the kernel of the
operator $ T (z)-T(\delta; |z|),\,z\neq u_0\,(u_0\neq 0)$ by
\begin{equation*}
 C [ (p^2+q^2)^{-1} +
|p|^{-\frac{1}{2}}(p^2+q^2)^{-1} +
|q|^{-\frac{1}{2}}(p^2+q^2)^{-1}+1 ]
\end{equation*}
 and hence
the operator $ T(z)-T(\delta; |z|),\,z\neq u_0$
 belongs to the Hilbert-Schmidt class for all
$z \leq 0.$ In combination with the continuity of the kernel of
the operator in  $z<0$ this  gives   the continuity of $
T(z)-T(\delta;|z|),\,z\neq u_0$ in  $z\leq 0.$ The details are omitted.
\end{proof}

Let us now recall some results from \cite{Sob}, which are
important in our work. \\ Let ${\bf S}_{{\bf r}}:L_2((0,{\bf
r})\times {\bf \sigma})\to L_2((0,{\bf r})\times {\bf \sigma}) $
be the integral operator with the kernel
\begin{equation*}
 S(y;t)=(2\pi)^{-2}\frac{l}
{\cosh y+st}
\end{equation*}
and
\begin{align*}
{\bf r}=1/2 | \log |z||,\,y=x-x',\,t=<\xi, \eta>,\,\xi, \eta \in
\S^2,\,l=\big( \frac{l^2_1}{l^2_1-l^2_2} \big)^{\frac{1}{2}},\,
 s=\frac{l_2} {l_1},
\end{align*}
${\bf \sigma}=L_2(\S^2),\, \S^2$ being the unit sphere in $\R^3.$

The coefficient in the asymptotics of $N(z)$ will be expressed by
means of the self adjoint integral operator $\hat{\bf
S}(\lambda),\,\lambda\in \R,$ in the space $L_2(\S^2)$ whose
kernel depends on the scalar product $t=<\xi,\eta>$ of the
arguments $\xi,\eta\in\S^2$ and has the form
\begin{equation*}
\hat {\bf S}(\lambda)=(2\pi)^{-1}l\frac{\sinh[\lambda(arccos st)]}
{(1-s^2t^2)^{\frac{1}{2}}\sinh (\pi\lambda)}.
\end{equation*}

For $\mu>0,$ define
$$
{U}(\mu)= (4\pi)^{-1} \int\limits_{-\infty}^{+\infty}
n(\mu,\hat{\bf S}(y))dy
$$
and set $\cU_0\equiv U(1).$

The following lemma can be proved in the same way as Theorem 4.5
in \cite{Sob}.
\begin{lemma}\label{lim} The following equality
$$ \lim\limits_{{\bf r}\to \infty} \frac{1}{2}{\bf
r}^{-1}n(\mu,{\bf S}_{\bf r})={U}(\mu) $$ holds.
\end{lemma}

The following theorem is basic for the proof of the asymptotics
\eqref{asym.K}.
\begin{theorem}\label{main} Let $u_0\neq 0$ (with $u_0$ as in ). The equality
$$ \lim\limits_{|z|\to 0} \frac{n(1,T(z))} {|log|z||}
=\lim\limits_{{\bf r}\to \infty} \frac{1}{2}{\bf r}^{-1}n(1,{\bf
S}_{\bf r}) $$ holds.
\end{theorem}

\begin{remark} Since $\cU(\cdot)$ is continuous in $\mu,$ according
to Lemma \ref{comp.pert} any perturbations of the operator $A_0(z)$
defined in Lemma \ref{comp.pert}, which is compact and continuous up
to $z=0$ do not contribute to the asymptotics \eqref{asym.K}. During
the proof of Theorem \ref{main} we use this fact without further
comments.
\end{remark}
{\bf Proof of Theorem \ref{main}.} Let $u_0\neq 0.$ As in Lemma
\ref{H-Sh}, it can be shown that $ T(z)-T(\delta; |z|),\,z\neq
u_0$ defines a compact operator continuous in $z\le 0$  and it
does not contribute to the asymptotics \eqref{asym.K}.

The space of functions  having support in $\hat U_\delta(0)$  is
an invariant subspace for the operator $T(\delta;|z|),\,z\neq
u_0.$

Let  $\hat T_0(\delta;|z|),\,z\neq u_0$ be the restriction of the
operator $T (\delta;|z|),\,z\neq u_0$ to the subspace $L_2(\hat
U_{\delta}(0)).$ One verifies  that the operator $\hat
T_0(\delta;|z|),\,z\neq u_0$ is unitary equivalent to the
following operator $T_0(\delta;|z|),\,z\neq u_0$ acting in
$L_2(\hat U_{\delta}(0))$ as
\begin{align*}
(T_0(\delta;|z|)f)(p)= \mathrm{d}_1 \int_{U_\delta(0)} \frac{
(n p^2+ 2|z|)^{-1/4} (n q^2+ 2|z|)^ {-1/4} } {l_1 p^2+
2l_2(p,q)+l_1 q^2 +2|z|} f(q)d q,
\end{align*}
where
\begin{align*}
 d_1= (2 \pi^2)^{-1}l_1^{{3}/{2}}.
\end{align*}

Here the equivalence is performed by the unitary dilation $${\bf
Y}:L_2( U_{\delta}(0))
 \to L_2(\hat U_{\delta}(0)),\quad
 ({\bf Y} f)(p)=f(U^{-\frac{1}{2}}p).
 $$

The operator $T_0(\delta;|z|),\,z\neq u_0$ is unitary equivalent
 to the integral operator $T_1(\delta;|z|):L_2(U_r(0)) \to L_2(U_r(0))$
 with the kernel
\begin{align*}
 \mathrm{d}_1 \frac{  (n p^2+ 2|z|)^{-1/4} (n  q^2+ 2|z|)^ {-1/4} } {l_1 p^2+
2l_2(p,q)+l_1 q^2 +2},
\end{align*}
where $U_r(0)=\{ p\in \R^3:|p|<r\},\,\,r=|z|^{-\frac{1}{2}}, \,z\neq u_0.$

The equivalence is performed by the unitary dilation
 $${\bf B}_r:L_2
(U_\delta(0)) \to L_2(U_r(0)),\quad
 ({\bf B}_r f)(p)=(\frac{r}{\delta})^{-3/2}f(\frac{\delta}
{r}p).$$ Further, we may replace $$(n p^2+2)^{-1/4},\, (n
q^2+2)^{-1/4} \quad \mbox{ and}\quad l_1 p^2+ 2l_2(p,q)+l_1 q^2
 +2$$
 by
$$(n p^2)^{-1/4}(1-\chi_1(p)),\,\, (n
q^2)^{-1/4}(1-\chi_1(q))
 \quad \mbox{ and}\quad
l_1 p^2+ 2l_2(p,q)+l_1 q^2 ,$$
  respectively, since the error will be
a Hilbert-Schmidt operator  continuous up to  $z=0$. Then we get
the integral operator $T_2(r)$ on $L_2(U_r(0) \setminus U_1(0))$  with the kernel
\begin{align*}
(n)^{-\frac{1}{2}} \mathrm{d}_1 \frac{|p|^{-1/2}| q|^{-1/2}} {l_1  p^2+
2l_2(p,q)+l_1 q^2}.
\end{align*}

By the dilation $${\bf M}:L_2(U_r(0) \setminus U_1(0))
\longrightarrow L_2((0,{\bf r})\times {\bf \sigma}),$$
 where
$(M\,f)(x,w)=e^{3x/2}f(e^{ x}w), x\in (0,{\bf r}),\, w \in
{\S}^2,$ one sees that the operator $T_2(r)$ is unitary equivalent
with the integral operator ${\bf S}_{{\bf r}}.$ \qed

{\bf Proof of (ii) of Theorem \ref{fin}} Let the conditions
 part $(ii)$ of Theorem \ref{fin} be fulfilled.

{\bf Case $u_0\neq 0.$} Similarly to \cite{Sob} we can show that
\begin{equation}\label{Slambda}
\cU_0={U}(1) \ge \frac{1}{4\pi} \int\limits_{-\infty}^{+\infty}
n(1,\hat{\bf S}^{(0)}(y))dy\ge\frac{1}{4\pi} mes\{y:\hat{S}^{(0)}(y)>1\},
\end{equation}
where $\hat{\bf S}^{(0)}(y)$  is the multiplication operator by
number
\begin{equation*}
\hat {S}^{(0)}(y)=l\frac{\sinh(y arcsin s)}
{sy\cosh \frac{\pi y}{2}}
\end{equation*}
in the subspace of the harmonics of degree zero.

The positivity of $\cU_0$ follows  from the facts that $l>1,\,\hat
{S}^{(0)}(0)>1$ and continuity of $\hat {S}^{(0)}(y).$ Taking into
account the inequality \eqref{Slambda} and Lemmas \ref{b-s},
\ref{main}, \ref{lim}, we complete the proof of $(ii)$ of Theorem
\ref{fin} in case $u_0\neq 0.$

{\bf Case $u_0=0.$}
 Since $n(1,T^{(2)}(z))\leq 1$ for any $z<0$ according to the
 Weyl inequality for all
$z<0$ and $\gamma \in (0,1)$ we obtain
$$
n(1+\gamma,T^{(1)}(z))\leq n(1,T(z))\leq
n(1-\gamma,T^{(1)}(z))+1.
$$

From the latter inequality we have
$$
{U}(1+\gamma)\leq \lim\limits_{|z|\to 0} \frac{n(1,T(z))}
{|log|z||}  \leq {U}(1-\gamma).
$$

After this remark from the continuity of the function $U(\cdot)$ it
follows
$$
\lim\limits_{|z|\to 0} \frac{n(1,T(z))} {|log|z||}  =
{U}(1)=\cU_0.
$$

The latter equality and Lemma \ref{b-s} completes the proof of
$(ii)$ of Theorem \ref{fin} in case $u_0=0.$ \qed

\appendix
\section{}

\begin{lemma}\label{examp}
Let
$$
u(p)= \varepsilon (p)+c,\, v(p)= \varepsilon (p),\,
w(p,q)=\varepsilon (p)+ \varepsilon (p+q)+ \varepsilon (q),
$$
where $c>0-$  is a real number and the function $\varepsilon (p)$
is defined by \eqref{eps.}
 Then Assumptions \ref{hypoth1}, \ref{hypoth2} and \ref{hypoth3} are fulfilled.
\end{lemma}
\begin{proof}
It is easy to see that Assumptions \ref{hypoth1} and \ref{hypoth2}
are fulfilled.

We prove that  Assumption \ref{hypoth3} is fulfilled. Since
$w(p,q)$ and $v(p)$ are even the function $\Lambda(p)$ is also
even.

Then we get
\begin{equation}\label{L.Even}
\Lambda(p)-\Lambda(0)= \frac{1}{4}\int\limits_{{\T}^3}
\frac{2w_0(t)-(w_p(t)+w_{-p}(t))}
{w_p(t)w_{-p}(t)w_0(t)}[w_p(t)+w_{-p}(t)]v^2(t)dt-
\end{equation}
$$
-\frac{1}{4}\int\limits_{{\T}^3} \frac{ [w_p(t)-w_{-p}(t)]^2}
{w_p(t)w_{-p}(t)w_0(t)}v^2(t)dt.
$$
From the equality
\begin{equation*}
w_0(t)-\frac{w_p(t)+w_p(-t)}{2}= \sum_{j=1}^{3}(\cos p_i-1)(1+\cos
t_i)
\end{equation*}
and \eqref{L.Even}  we get for all nonzero $p\in \T^3$ the
inequality
\begin{equation*}
\Lambda(p)-\Lambda(0)= \int\limits_{{\T}^3} \frac{
\sum_{j=1}^{3}(\cos p_i-1)(1+\cos t_i)
v^2(t)dt}{w_p(t)w_p(-t)w_0(t)}-\frac{1}{4}\int\limits_{{\T}^3}
\frac{ [w_p(t)-w_{-p}(t)]^2} {w_p(t)w_{-p}(t)w_0(t)}v^2(t)dt<0,
\end{equation*}
that is, part  $(i)$ of Assumption \ref{hypoth3} holds.

Since for any $p,q\in \T^3,p\neq 0$ the inequality $w_p(q)>0$
holds for any nonzero $p\in \T^3$ the integrals
\begin{equation*}\label{L.Def}
\int\limits_{{\T}^3} \frac{ \frac{\partial^2 } {\partial p_i
\partial p_j}w_p(t) v^2(t)dt}{(w_p(t))^2}
\end{equation*}
and
\begin{equation*}\label{L.Def2}
2\int\limits_{{\T}^3} \frac{ \frac{\partial } {\partial p_i}
w_p(t) \frac{\partial } {\partial p_j} w_p(t)
v^2(t)dt}{(w_p(t))^3},\quad i,j=1,2,3
\end{equation*}
are finite and the finiteness of these integrals at the point
$p=0$ follows the fact that $v(0)=0.$ After this remark we can
define bounded continuous functions on $\T^3,$ which will be
denotes by $\lambda_{ij}^{(1)}(p)$ and $\lambda_{ij}^{(2)}(p),$
respectively.

Then the function $\Lambda(p)$ is a twice continuously differentiable function
on $\T^3$ and
\begin{equation*}\label{L.Def3}
\frac{\partial^2 \Lambda(p)} {\partial p_i
\partial p_j}=-\lambda_{ij}^{(1)}(p)+\lambda_{ij}^{(2)}(p),
\quad i,j=1,2,3.
\end{equation*}

Since
\begin{align*}\label{deff}
&  \frac{\partial } {\partial p_i }w_0(t)
=\sin t_i,\\
&  \frac{\partial^2 } {\partial p_i\partial p_i
}w_0(t) =1+\cos t_i,\nonumber\\
& \frac{\partial^2 } {\partial p_i\partial p_j }w_0(t)=0,\quad
i\neq j,\,\,i,j=1,2,3\nonumber
\end{align*}
we get
\begin{equation*}\label{L.Def4}
\frac{\partial^2 \Lambda(0)} {\partial p_i
\partial p_i}=-
2\int\limits_{{\T}^3} \frac{ \sum_{s=1,s\neq i}^3 (1-\cos
t_s)(1+\cos t_i) v^2(t)dt}{(w_0(t))^3} ,
\end{equation*}
\begin{equation*}\label{L.Def5}
\frac{\partial^2 \Lambda(0)} {\partial p_i
\partial p_j}=2
\int\limits_{{\T}^3} \frac{ \sin t_i \sin t_j
v^2(t)dt}{(w_0(t))^3},\quad i\neq j,\, i,j=1,2,3.
\end{equation*}

Since the function $v$ is even on $\T^3$ from the latter two
equalities we get
\begin{equation*}
\frac{\partial^2 \Lambda(0)} {\partial p_i
\partial p_i}<0,
\frac{\partial^2 \Lambda(0)} {\partial p_i
\partial p_j}=0,\,\,
i\neq j,\,\,i,j=1,2,3.
\end{equation*}

Using these facts, one may verify that the matrix of the second
order partial derivatives of the function $ \Lambda(p)$ at the
point $p=0$ is negative definite. Thus the function $\Lambda(p)$
has a non-degenerated
 maximum at the point $p=0.$
\end{proof}

\begin{lemma}\label{int finite} Let $v$ be an arbitrary analytic function on $\T^3$ and
$$ w_p(q)=l_1\varepsilon (p)+ l_2\varepsilon (p+q)+ l_1\varepsilon (q),
$$
where $l_j>0,j=1,2,\, l_1 \neq l_2$ and the function
$\varepsilon(p)$ is defined by \eqref{eps}.
 Then for any $p\in \T^3$ and
$\psi\in C(\T^3)$ the integral
$$
\int\limits_{{\T}^3} \frac{v(t)\psi(t)dt}{w_p(t)-m_w(p)}
$$
is finite.
\end{lemma}
\begin{proof}
The function $w_p(q)$ can be rewritten in the form
\begin{equation}\label{echiladigan}
 w_p(q)=\varepsilon_1 (p)+3(l_1+l_2)-\sum_{i=1}^3
 (a(p_j)\cos q_j+c(p_j)\sin p_j),
\end{equation}
where the coefficients $a(p_j)$ and $c(p_j)$ are given by
\begin{equation*}
a(p_j)=l_2 \cos p_j+ l_1, c(p_j)=-l_2 \sin p_j.
\end{equation*}

The equality \eqref{echiladigan} implies the following
representation for $w_p(q)$
\begin{equation}\label{represent}
w_p(q)=\varepsilon_1 (p)+3(l_1+l_2)- \sum_{i=1}^3r(p_i)(\cos
q_i-q_0(p_i)).
\end{equation}
where
$$
r(p_i)=\sqrt{a^2(p_i)+c^2(p_i)},\quad
q_0(p_i)=\arcsin\frac{c(p_i)}{r(p_i)},\quad p_i \in (-\pi,\pi].$$

Therefore
\begin{equation}\label{rep mw}
m_w(p)=\min_{q\in \T^3}w_p(q)=\varepsilon_1 (p)+3(l_1+l_2)-
\sum_{i=1}^3r(p_i).
\end{equation}

From \eqref{represent} and \eqref{rep mw}  we have
$$
w_p(q)-m_w(p)=\sum_{i=1}^3r(p_i)(1-\cos (q_i-q_0(p_i)).
$$
Since $l_j>0,j=1,2,\,l_1 \neq l_2$ for any $p\in \T^3$ the
function $w_p(q)-m_w(p)$  has a unique non-degenerate minimum at
the point $q=q_0(p)=(q_0(p_1),q_0(p_2),q_0(p_3)),$ therefore for
any $p\in \T^3$ the integral
$$\int\limits_{{\T}^3} \frac{v(t)\psi(t)dt}{w_p(t)-m_w(p)}$$
is finite.
\end{proof}

\begin{lemma}\label{minimum}
Let Assumption \ref{hypoth1} be fulfilled. Then there exists a $ {
\delta } $-neighborhood $U_ {\delta } (0)\subset \T^3$ of the point
$p=0$ and an analytic function $q_0 (p)$ defined on $U_ {\delta }
(0) $
  such
  that:\\
(i)   for any $p { \in } U_ { \delta }(0)$  the point
  $q_0(p) $ is
a unique
  non-degenerate minimum  of the function
$w_p(q)$ and
\begin{equation*}
q_0(p)=-\frac{l_2}{l_1}p+O(|p|^3)\,\,as\,\,p \to 0.
\end{equation*}\\
(ii) The function $m_w(p)= w_{p}(q_0(p))$ is analytic in $U_ {\delta
} (  0 ) $ and has the asymptotic form
 \begin{equation}\label{min.raz}
m_w(p)=\frac{l^2_1-l^2_2}{2l_2}(Wp,p)+O(|p|^4) \quad
\mbox{as}\quad p \to 0.
\end{equation}
\end{lemma}
\begin{proof}
(i) By Assumption \ref{hypoth1} we obtain $ w_0(q)>w_0(0),\,q
\neq 0 $ and
 $$
\left( \frac{\partial^2 w_0(0)}{\partial q^{(i)}
\partial q^{(j)}} \right)_{i,j=1}^3= l_1 W.
$$

Since $W$ is a positive matrix  the function $ w_0(q)$
 has a unique non-degenerate  minimum at  $q=0,$
 the gradient
  $ { \bigtriangledown } w_0(q) $
   is equal to  zero at the point $q=0.$

Now we apply the implicit function  theorem to the equation
$$
{\bigtriangledown } w_p(q)=0,\,\,p,q\in {\T}^3.
$$
Then there exists a $ { \delta } $-neighborhood  $U_ { \delta } (0
) $ of the point $p=0$ and a vector function $q_0 (p)$  defined
and analytic in $U_ { \delta }(0) $ and for all $p { \in } U_ {
\delta } ( 0 ) $ the identity $ { \bigtriangledown } w _p( q_0(p))
 \equiv 0 $ holds.

Denote by $B(p) $ the matrix of the second order partial
derivatives of the function $ w_p( q) $
 at the point
$ q_0(p)$. The matrix $B(0)=l_1 W$ is positive definite and  $B
(p)$ is continuous in $U_ { \delta } ( 0 )$ and hence  for any $p
\in U_{\delta} ( 0 )$
 the matrix $B (p)$ is positive definite. Thus
$ q_0(p),\,p \in U_ { \delta } ( 0 )$ is the unique non-degenerate
minimum point
 of $w_p(q).$

The    non-degenerate  minimum point $q_0(p)$ is an odd function
in $p \in U_\delta(0).$

Indeed, since $w(p,q)$ is even with respect $(p,q)$ we get $w_p( -q)=w_{-p}( q)$, and
we obtain $$ w_{-p}(-q_0(p))= m_w(p)=m_w(-p)= w_{-p}(q_0(-p)). $$

Since for all $p \in U_\delta(0)$ the point $q_0(p)$ is the unique
non-degenerate  minimum of $w_{p}(q)$ we have $$ q_0(-p)=-q_0(p).
$$

By Assumption \ref{hypoth1} and the Taylor expansion
 we get
 \begin{equation}\label{taylor2}
w_p(q -\frac{l_2}{l_1}p )=
 \frac{l_1}{2}(Wq,q)+
\frac{l^2_1-l^2_2}{2l_1}(Wp,p)+
 O(|q|^4+|p|^4)\,\,\mbox{as}\,\,q,p
\to 0.
\end{equation}

Since $w_p(q_0(p))=\min\limits_{q\in {\T}^3}
 w_p  (q ) \leq
w_p(-\frac{l_2}{l_1}p)$ and $q_0(p)$ is odd
 we have
$$
\frac{l_1}{2}\big(W^{\frac{1}{2}}(q_0(p)+\frac{l_2}{l_1}p ) \big
)^2+ \frac{l^2_1-l^2_2}{2l_1}(Wp,p)+
 O(|p|^4)\leq
 \frac{l^2_1-l^2_2}{2l_1}(Wp,p)+
 O(|p|^4)
 \,\,\mbox{as}\,\,p
\to 0,
 $$
 that is,
$$
l_1\big(W^{\frac{1}{2}}(q_0(p)+\frac{l_2}{l_1}p ) \big )^2 \leq
O(|p|^4) \,\,\mbox{as}\,\,p \in U_\delta(0).
$$

This inequality is not valid if $q_0(p)$ has the asymptotics
$q_0(p) + \frac{l_2}{l_1}p=O(|p|)$ as $p \to 0.$ Since $q_0(p)$ is
an odd analytic function, we have
$q_0(p)+\frac{l_2}{l_1}p=O(|p|^3)\,\,as\,\,p \to 0.$

(ii) Since the functions $w(p,q),\,p,q \in \T^3$ and $q_0(p),\,
p\in U_{\delta}(0)$
 are analytic we have that the function
$m_w(p)=w_p(q_0(p))$ is also analytic on $p \in U_{\delta}(0) $.

By $q_0(p)=-\frac{l_2}{l_1}p+O(|p|^3),\,p \to 0$ and
\eqref{taylor2} we get the asymptotics \eqref{min.raz}.
\end{proof}
\begin{lemma}\label{U.ineq}
There exist numbers $C_1, C_2,C_3>0$ and $\delta>0$ such that the
following inequalities
\begin{align*}
&(i)\quad C_1 (|p|^2+|q|^2)\leq w(p,q) \leq C_2 (|p|^2+|q|^2)\quad
\mbox{for all} \quad p, q\in U_\delta(0),\\
 &(ii)\quad w(p,q) \geq C_3\quad \mbox{for all} \quad(p, q)\notin
U_\delta(0)\times U_\delta(0)
\end{align*}
 hold.
\end{lemma}
\begin{proof}
 By Assumption \ref{hypoth1} the point $(0,0)\in (\T^3)^2$ is
unique non-degenerated minimum of $w(p,q).$ Then by \eqref{taylor2}
there exist positive  numbers $C_1,C_2,C_3$ and a
$\delta-$neighborhood of $p=0\in \T^3$ so that $(i)$ and $(ii)$ hold
true.
\end{proof}

{\bf Acknowledgement}

This work was supported by DFG 436 USB 113/4 and DFG 436 USB 113/6
projects and the Fundamental Science Foundation of Uzbekistan. The
last two named authors gratefully acknowledged the hospitality of
the Institute of Applied Mathematics and of the IZKS of the
University Bonn.


\begin{thebibliography}{99}

\bibitem{AGH}  S. Albeverio, F. Gesztesy , and R. H{\o}egh-Krohn,
The low energy expansion in non-relativistic scattering theory, Ann.
Inst. H. Poincar\'e Sect. A (N.S.)  37 (1982), 1--28.

\bibitem{AHW}  { S. Albeverio, R. H{\o }egh-Krohn}, and {\sc T. T.
Wu}, {A class of exactly solvable three--body quantum mechanical
problems and universal low energy behavior}, Phys. Lett. A {83}
(1971), 105-109.

\bibitem{ALzMahp04}  {S.~ Albeverio, S.~N.~Lakaev, and
 Z.~I.~Muminov}, { Schr\"{o}dinger operators on lattices. The Efimov effect and
discrete spectrum asymptotics.} Ann. Henri Poincar\'{e}. { 5},
(2004) 743--772.

 \bibitem{ALaams97}  {Zh.~I. ~Abdullaev, S.~N. ~Lakaev}, {On the
spectral properties of the matrix-valued Friedrichs model.
Many-particle Hamiltonians: spectra and scattering,}  1--37, Adv.
Soviet Math., 5, Amer. Math. Soc., Providence, RI, 1991.

\bibitem{ALtmf03}  { ~J. ~I. ~Abdullaev,~S.~N. ~Lakaev}, { Asymptotics of the Discrete Spectrum
of the Three-Particle Schr"dinger Difference Operator on a Lattice},
Theoretical and Mathematical Physics 136 (2003), No.2, 1096-1109;


\bibitem{AmNo}  { ~R.~D. ~Amado and ~J.~V. ~Noble,} { Efimov effect; A new pathology of
three-particle Systems,} II.Phys.Lett.B.35.No.1,25-27,(1971);
Phys.Lett.D.5. No.8, (1972), 1992-2002.

\bibitem{DFT} {~G. ~F. ~Dell'Antonio, ~R. ~Figari, ~A. ~Teta,} { Hamiltonians for systems of $N$ particles interacting
through point interactions,} Ann. Inst. H. Poincaré Phys. Théor. 60
(1994), no. 3, 253--290.

\bibitem{FIC} { ~P.~A.~Faria da Veiga, ~L.~Ioriatti, and
M.~O'Carroll}, {  Energy-momentum spectrum of some two-particle
lattice Schr\"odinger Hamiltonians,} Phys. Rev. E (3)
 {66}, (2002) \, 016130, 9 pp.

\bibitem{GrSc} { ~G.~M.~Graf and ~D.~Schenker},
{ $2$-magnon scattering in the Heisenberg model}, Ann. Inst. H.
Poincar\'e Phys. Th\'eor. {67} (1997),  91--107.

\bibitem{Efi}  {~V. ~Efimov}, { Energy levels of three resonantly interacting
particles}, Nucl. Phys. A { 210} (1973), 157--158.

\bibitem{FaMe}  { ~L. ~D. ~Faddeev} and { ~S. ~P. ~Merkuriev}, { Quantum
scattering theory for several particle systems,} Kluwer Academic
Publishers, 1993.

\bibitem{Frie}  { ~K. ~O. ~Friedrichs}, {Perturbation of spectra in
Hilbert space,} 1965, American Mathematical Society providence,
Rhode Island.

\bibitem{KM} ~Yu.~G. ~Kondratiev  and R.~A.~Minlos ,
 One-particle subspaces in the stochastic $XY$ model,
  J. Statist. Phys. {87} (1997), 613--642.



\bibitem{Ltmf91}  {~S.~N. ~Lakaev}, { On an infinite number of three-particle bound
states of a system of quantum lattice particles,} Theor.and
Math.Phys.{89} (1991),No.1, 1079--1086.

\bibitem{LtsP86}  { ~S.~N. ~Lakaev }, Some spectral properties of the generalized
Friedrichs model, (Russian) Trudy Sem. Petrovsk. No. 11 (1986),
210--238, 246, 248; translation in J. Soviet Math. 45 (1989), no. 6,
1540--1565

\bibitem{Ltmf92}  { ~S.~N. ~Lakaev,} { Bound states and resonances fo the N-particle
discrete Schr\"{o}dinger operator,} Theor.Math.Phys.91
(1992),No.1,362-372.

\bibitem{Lfa93}  {~S.~N. ~Lakaev}, {The Efimov's Effect of a system of Three
Identical Quantum lattice Particles,} Funkcionalnii analiz i ego
priloj. , {27} (1993), No.3, pp.15-28, translation in Funct.
Anal.Appl.

\bibitem{LAfa99}  {~S.~N. ~Lakaev and ~J.~I. ~Abdullaev}, { The spectral properties of
the three-particle difference Schr\"{o}dinger operator,} Funct.Anal.
Appl. {33} (1999), No. 2, 84-88.

\bibitem{LRfa03}  {~S.~N. ~Lakaev and ~T.~Kh. ~Rasulov}, {Efimov's Effect in a Model of
Perturbation Theory of the Essential Spectrum,} Funct.Anal. Appl.
{37} (2003), No. 1, 69-71.

\bibitem{LRmn03}  {~S.~N. ~Lakaev and ~T.~Kh. ~Rasulov}, A Model in the Theory of Perturbations
of the Essential Spectrum of Multiparticle Operators, Mathematical
Notes 73 (2003),No 3, 521-528;


\bibitem{Mat}  {~D.~C. ~Mattis}, The few-body problem on lattice, Rev.Modern
Phys. { 58} (1986), No. 2, 361-379


\bibitem{MS} ~R.~A. Minlos  and ~Y.~M. Suhov,
On the spectrum of the generator of an infinite system of
interacting diffusions, Comm. Math. Phys. {206} (1999),
  463--489.

\bibitem{MiSp} {~R. ~Minlos and ~H. ~Spohn}, The three-body problem in radioaction
desay: the case of one atom and at most two photons,
Amer.Math.Soc.Transl.(2){ 177} (1996),159-193

\bibitem{Mog91}{ ~A.~I. ~Mogilner}, Hamiltonians of solid state physics at
few-particle discrete Schrodinger operators: problems and results,
Advances in Sov. Math., {5} (1991),139-194

\bibitem{OvSi}  { ~Yu.~N.~Ovchinnikov and ~I.~M. ~Sigal}, Number of bound states of
three-particle systems and Efimov's effect, Ann. Physics, {123}
(1989),274-295

\bibitem{rauch}  {~J. ~Rauch}, Perturbation theory for
eigenvalues and resonances of Schr\"odinger Hamiltonians,  J. Funct.
Anal.  35  (1980), no. 3, 304--315.

\bibitem{RSIII}  { ~M. ~Reed} and { ~B. ~Simon,} Methods of modern mathematical
physics. III: Scattering teory, Academic Press, N.Y., 1979.

\bibitem{RSIV}  { ~M. ~Reed} and { ~B. ~Simon,} Methods of modern mathematical
physics. IV: Analysis of Operators, Academic Press, N.Y., 1979.

\bibitem{Sob}  {~A.~V. ~Sobolev}, The Efimov effect. Discrete spectrum
asymptotics, Commun. Math. Phys. {156} (1993), 127--168.

\bibitem{Tam91}  { ~H.~ Tamura}, The Efimov effect of three-body Schr\"{o}dinger operator, J.
Funct. Anal. {95} (1991), 433--459.

\bibitem{Tam94}  { ~H.~ Tamura}, Asymptotics for the number of negative eigenvalues of three-body
Schrödinger operators with Efimov effect. Spectral and scattering
theory and applications, 311--322, Adv. Stud. Pure Math., 23,
Math. Soc. Japan, Tokyo, 1994.

\bibitem{Wang}  { ~X.~P~. Wang},  On the existence of the $N-$ body Efimov effect, J.
Funct. Anal. {95} (1991), 433--459.

\bibitem{Yaf74}  { ~D.~ R.~ Yafaev}, On the theory of the discrete spectrum of
the three-particle Schr\"{o}dinger operator, Math. USSR-Sb. { 23}
(1974), 535--559.

\bibitem{Yaf00}   ~D.~R.~Yafaev,
 { Scattering theory: Some old and new problems},
 Lecture Notes in Mathematics, 1735.
 Springer-Verlag, Berlin, 2000, 169 pp.



\bibitem{MiZh} { ~Yu. ~Zhukov, ~R. ~Minlos}, The spectrum and scattering in the
"spin-boson" model with at most three photons. (Russian) Teoret.
Mat. Fiz. 103 (1995), no. 1, 63--81; translation in Theoret. and
Math. Phys. 103 (1995), no. 1, 398--411


\bibitem{Zh}  ~E.~A. ~Zhizhina,
 Two-particle spectrum of the generator for stochastic model
of planar rotators at high temperatures,
 J. Statist. Phys.  {91} (1998),  343--368.
\end{thebibliography}
\end{document}